\documentclass[conference]{IEEEtran}
\IEEEoverridecommandlockouts
\usepackage{cite}
\usepackage{amsmath,amssymb,amsfonts}
\usepackage{algorithmic}
\usepackage{graphicx}
\usepackage{textcomp}
\usepackage{xcolor}
\usepackage{url}
\usepackage{array}
\usepackage{comment}
\usepackage{subcaption}
\usepackage{hyperref}
\def\BibTeX{{\rm B\kern-.05em{\sc i\kern-.025em b}\kern-.08em
    T\kern-.1667em\lower.7ex\hbox{E}\kern-.125emX}}

\providecommand{\KL}[1]{\textcolor{purple}{[{\bf #1}]}}

\providecommand{\NZ}[1]{\textcolor{olive}{[{\bf #1}]}}

\begin{document}

\title{A Multi-Stream Fusion Approach with One-Class Learning for Audio-Visual Deepfake Detection\\
\thanks{This work is supported in part by a New York State Center of Excellence in Data Science award, National Institute of Justice (NIJ) Graduate Research Fellowship Award 15PNIJ-23-GG-01933-RESS, and synergistic activities funded by National Science Foundation (NSF) grant DGE-1922591.}
}

\author{


\IEEEauthorblockN{Kyungbok Lee$^1$, You Zhang$^2$, Zhiyao Duan$^2$}
\IEEEauthorblockA{
\textit{$^1$Department of Computer Science, $^2$Department of Electrical and Computer Engineering} \\
University of Rochester,
Rochester, NY, USA \\
klee109@u.rochester.edu, \{you.zhang, zhiyao.duan\}@rochester.edu}

}

\maketitle
\begin{abstract}
This paper addresses the challenge of developing a robust audio-visual deepfake detection model. In practical use cases, new generation algorithms are continually emerging, and these algorithms are not encountered during the development of detection methods. This calls for the generalization ability of the method. Additionally, to ensure the credibility of detection methods, it is beneficial for the model to interpret which cues from the video indicate it is fake. Motivated by these considerations, we then propose a multi-stream fusion approach with one-class learning as a representation-level regularization technique. We study the generalization problem of audio-visual deepfake detection by creating a new benchmark by extending and re-splitting the existing FakeAVCeleb dataset. The benchmark contains four categories of fake videos (Real Audio-Fake Visual, Fake Audio-Fake Visual, Fake Audio-Real Visual, and Unsynchronized videos). 
The experimental results demonstrate that our approach surpasses the previous models by a large margin.
Furthermore, our proposed framework offers interpretability, indicating which modality the model identifies as more likely to be fake. The source code is released at \textcolor{magenta}{\url{https://github.com/bok-bok/MSOC}}.

\end{abstract}    
\section{Introduction}
\label{sec:intro}


Recent advancements in deep learning, including Stable Diffusion \cite{rombach2022stablediff} and Sora, have enabled the generation of highly realistic images and audio, collectively referred to as deepfakes. The availability of numerous easy-to-use tools for generating deepfake videos significantly increases the chance of misuse of those media, as even non-experts can now create convincing fake content with minimal effort. This emphasizes the urgency for developing robust detection mechanisms to mitigate the risks associated with deepfakes.

Videos, particularly those featuring a person speaking, have become a significant medium for disseminating deepfake information. Detecting these deepfakes requires joint consideration of both audio and visual modalities. The speech could be generated from text-to-speech \cite{tan2021survey} and voice conversion algorithms \cite{desai2009voiceconversion, sisman2020overview}, and the videos are either face-swap~\cite{Korshunova2017faceswap} from an original video or further rendered from speech and a still image~\cite{sheng2024deep}. 

Additionally, while synchronization might be disrupted by modifying audio or visual modality, the generated modality can still be seamlessly synchronized with its corresponding counter modalities using lip-sync technologies \cite{prajwal2020lipsync, guan2023stylelipsync}. This ensures the creation of highly realistic fake videos.
This underscores the need for researchers to develop audio-visual deepfake detection mechanisms that surpass the capabilities of unimodal deepfake detection approaches.

Recent research focuses mainly on the fusion of features of both modalities to improve the detection performance on audio-visual deepfake datasets \cite{zou2024crossmrdf, muppalla2023integrating, zhou2021joint}. By leveraging the complementary nature of audio and visual data, these approaches effectively improve their accuracy in identifying manipulated content.
However, two issues are not well explored: First, the existing deep learning models may overfit to the specific fake generation methods present in the training data, leading to poor generalization when confronted with unseen deepfake generation algorithms in real-world scenarios. This could be attributed to the existing dataset design~\cite{khalid2021fakeavceleb, dolhansky2020dfdc} that does not benchmark the generalization ability for the models. This overfitting issue would limit the practical applicability of these models, as they fail to adapt to the rapidly evolving landscape of deepfake techniques. Second, existing approaches lack the ability to identify the modality source of a detected deepfake. This limitation arises because these systems are trained and tested using only the final audio-visual labels, without incorporating meta-information about the individual modalities.
A model able to tell which modality is fake would enhance the interpretability and credibility in practice.

In this work, we propose a novel framework \textbf{M}ulti-\textbf{S}tream Fusion Approach with \textbf{O}ne-\textbf{C}lass Learning (MSOC) to tackle audio-visual deepfake detection, enhancing the generalization ability and interoperability. We extend the one-class learning approach, previously proposed in uni-modal contexts, to the audio-visual setting. We validate the generalization ability by resplitting the FakeAVCeleb~\cite{khalid2021fakeavceleb} dataset and separating the unseen algorithms into the test set. We curated four test sets (RAFV, FAFV, FARV, Unsynced) 
that cover all kinds of fake categories.
We will make the dataset splits and model implementation publicly available upon the publication of this paper.
Our contributions are summarized as:
\begin{enumerate}  
\vspace{-0.2pt}
    \item Extending \textbf{one class learning} from uni-modal to audio-visual deepfake detection;
    \item A multi-stream framework with audio-visual (AV), audio (A), and visual (V) branches;
    \item A curated dataset for evaluating performance on unseen generation methods based on FakeAVCeleb. 
\end{enumerate}

\section{Related Work}
\label{sec:related}

\subsection{Audio-Visual Deepfake Detection}
At the early stage of deepfake detection research, many studies focused on uni-modal detection models that use only audio~\cite{Zhang2021OC, lu2024onedis} or visual~\cite{hu2022finfervideo} as input. However, uni-modal models are inherently limited to a single modality and cannot completely detect emerging deepfake videos that both audio and visual can be generated. To address this problem, recent research has started to focus on developing audio-visual deepfake detection models.

Initially, many studies have focused on explicit synchronization issues between audio and visual modalities in deepfakes.
Shahzad et al. \cite{shahzad2022lip} argue that altering either audio or visual can desynchronize speech and lip movements in videos. In addition, researchers have investigated the representation-level inconsistency due to single-modality manipulations. The modality dissonance score was introduced in \cite{chugh2020dissonance} to quantify the dissimilarities between the modality features. However, these methods may struggle to detect deepfakes where both audio and video are both generated in a more consistent way, such as text-to-speech followed by lip synch~\cite{khalid2021fakeavceleb}. 

Several studies also develop audio-visual representations by integrating features from uni-modal feature extractors and mapping them to audio-visual targets~\cite{zhou2021joint, yang2023avoiddf}. However, 
recent studies~\cite{muppalla2023integrating, zou2024crossmrdf} claim that using only multimodal labels can misinform the data from the unimodal feature extractor during joint training.
Ensemble models have also been studied~\cite{hashmi2023avtenet, hashmi2022multimodal, asha2022novel}. They combine models for audio, visual, and audio-visual data and leverage the strengths of each modality-specific model to enhance overall detection accuracy.

\subsection{Video Deepfake Detection Datasets}

Existing methods are typically benchmarked on datasets such as FakeAVCeleb~\cite{khalid2021fakeavceleb} and DFDC~\cite{dolhansky2020dfdc}. However, these datasets are limited in their ability to benchmark generalization since the test sets often contain the same deepfake generation algorithms as the training sets. Additionally, there is a greater variety of visual deepfake generation methods compared to audio modalities. In terms of attribute labeling, the FakeAVCeleb dataset attempts to present different categories of fakes, but the FARV category includes not only fake audio but also unsynchronized audio. This makes it difficult for methods to learn fake audio cues, since they are confounded with synchronization cues. Our study proposes extended datasets and new partitions to address these issues.





\subsection{One-Class Learning For Deepfake Detection}
Binary classification models work well in deepfake detection when the test data share similar distributions with the training data. However, since deepfake generation techniques are rapidly developing, deepfake attacks in practice are often unseen during training of deepfake detection models, and these binary classification models show significantly degraded performance on unseen attacks \cite{Ivanovska2021dis_oc}. 
To address this issue, Zhang et al. \cite{Zhang2021OC} proposed the idea of one-class learning for speech deepfake detection. The idea was to use a so-called One-Class Softmax (OC-Softmax) loss to guide the neural network to learn an embedding space where bonafide speech utterances are clustered together while fake speech utterances are pushed away from this cluster during training:
\begin{equation}
\mathcal{L}_{OCS} = \frac{1}{N} \sum_{i=1}^{N} \log \left( 1 + e^{\alpha \left( m_{y_i} - \mathbf{\hat{w}}^T \mathbf{\hat{x}}_i \right) (-1)^{y_i}} \right),
\label{eq:oc}
\end{equation}
where $\mathbf{\hat{w}}$ (center) is the normalized weight vector for the target class; $\mathbf{\hat{x}}_i$ is the normalized embedding vector of the $i$-th sample; $m_{0}$ and $m_{1}$ are margins for the real and fake classes, respectively. $N$ is the number of samples in a mini-batch, and $\alpha$ is a scale factor. For each utterance, the cosine similarity between the feature embedding and the weight vector, $\mathbf{\hat{w}}^T \mathbf{\hat{x}}_i$, is called the OC score, a value between -1 and 1.


Since then, many works on speech anti-spoofing have adopted the idea of one-class learning~\cite{ding2023samo, lu2024onedis, kim2024one}. The results show that models trained with one-class learning can effectively identify fakes as deviations from the learned bonafide embedding cluster for speech. 
Despite these advantages, the generalizability of one-class learning for audio-visual deepfake detection has not been thoroughly studied due to dataset limitations. This study addresses this gap by re-splitting the FakeAVCeleb dataset \cite{khalid2021fakeavceleb} and analyzing the effectiveness of one-class learning in audio-visual deepfake detection.

\section{Method}
\begin{figure*}[h!]
  \centering
  \includegraphics[width=\textwidth]{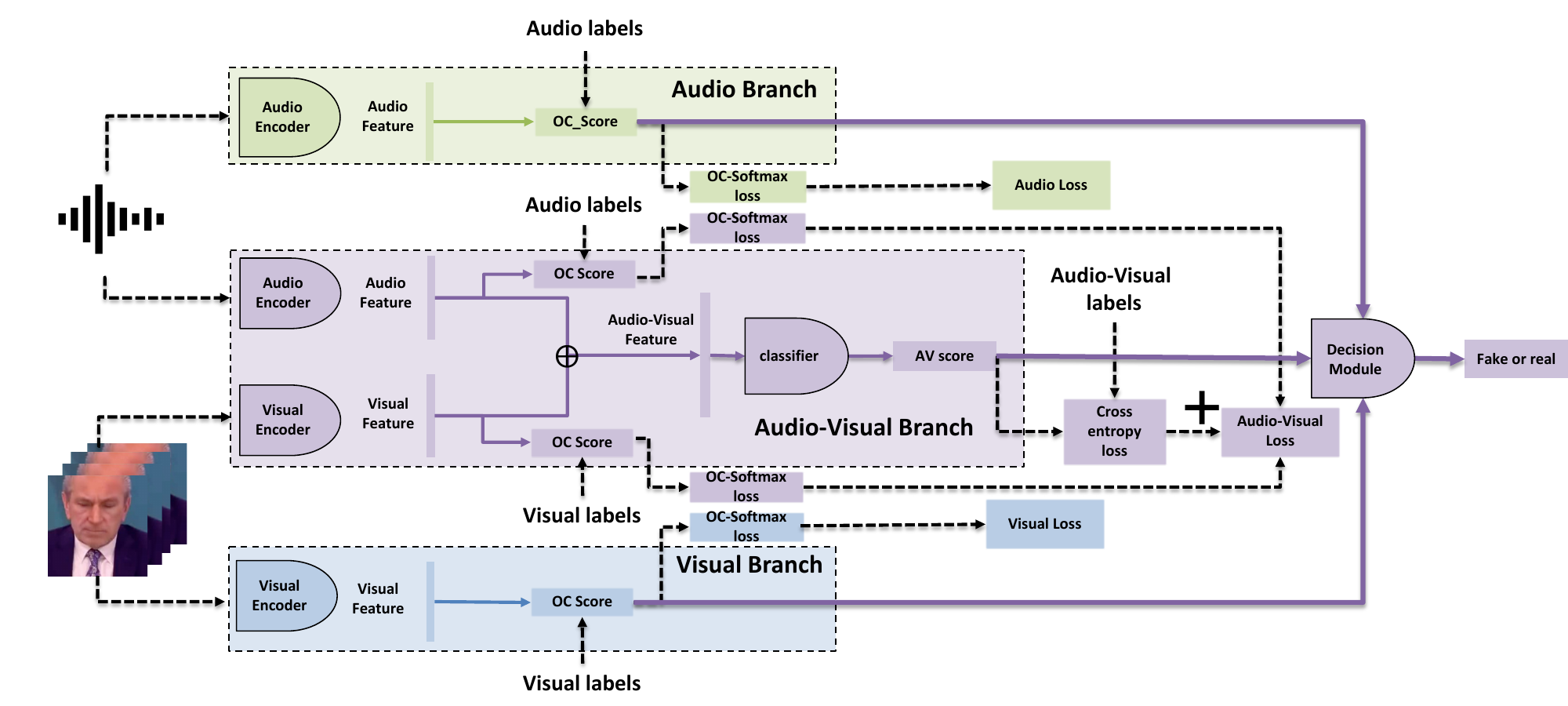}
  \caption{Overview of Multi-Stream Architecture. In the figure, black dashed lines represent the training process, and solid purple lines represent the inference process. $\bigoplus$ symbol represents feature concatenation. \textbf{+} symbol means addition.
  }
  
  \label{fig:model_structure}
\end{figure*}


We propose a \textbf{M}ulti-\textbf{S}tream Fusion Approach with \textbf{O}ne-\textbf{C}lass learning (\textbf{MSOC}) for audio-visual deepfake detection. This architecture consists of the audio, visual, and audio-visual branches, which are independently trained using labels specific to their respective modalities. The training of these branches also leverages the OC-Softmax loss to improve their generalization ability to unseen deepfake generation methods. During inference, score fusion is used to integrate the decisions made by the three branches to arrive at the final classification decision. In this section, we describe architecture, training, and inference in detail.

\subsection{Audio Model}
As displayed in the green part of Fig.~\ref{fig:model_structure}, the audio branch includes an audio feature extractor trained using an OC-Softmax loss with ground truth labels specific only to the audio modality. The audio branch compacts real data representations of audio modality and spreads fake ones in the embedding space. 

We utilize ResNet~\cite{he2016deep} as our audio feature extractor. The model processes 13-d Mel-Frequency Cepstral Coefficients (MFCC) vectors at a frame rate of 100 frames per second, which is 4 times the visual frame rate. 
The audio feature extractor then produces the audio embeddings with a dimensionality of 128. 

The audio model is trained with \(\mathcal{L}_{AOC}^{audio}\) which is the OC-Softmax losses computed with audio features of the audio branches using audio labels.

\subsection{Visual Model}
As shown in the blue part of Fig. \ref{fig:model_structure}, the visual branch consists of a visual feature extractor trained with an OC-Softmax loss, taking ground-truth labels regarding the visual modality only.
The visual branch tries to learn a visual embedding space where real data features are clustered while fake data features are pushed away from the cluster. 

We employed ResNet\cite{he2016deep} and SCNet\cite{liu2020scnet} with STIL (Spatiotemporal Inconsistency Learning) block\cite{gu2021spatiotemporal} as the visual feature extractor, which takes $T$ frames $v \in \mathbb{R}^{T \times 3 \times 100 \times 100}$, where 100 denotes the height and width of each frame, and 3 represents the RGB color channels. Then, the model returns embeddings $V \in \mathbb{R}^{T \times d}$, where the dimensionality $d$ is 128 for ResNet and 512 for SCNet-STIL.

\textbf{ResNet.}
The ResNet-based visual feature extractor consists of a 3D convolutional layer, ResNet blocks, and a temporal convolutional block. It captures the features of frames.

\textbf{SCNet-STIL.}
SCNet\cite{liu2020scnet} is a 2D Convolutional Neural Network. It features a self-calibration mechanism that broadens the receptive fields of its convolutional layers through internal communication \cite{liu2020scnet}. 
The SCNet-STIL is SCNet with STIL blocks designed to capture Spatio-Temporal Inconsistency \cite{gu2021spatiotemporal}. The STIL block is flexible and can be implemented in any 2D-CNN architecture.

The visual model is trained with \(\mathcal{L}_{VOC}^{visual}\), which is the OC-Softmax losses computed with visual features from the visual branch using visual labels.

\subsection{Audio-Visual Model}
As shown in the purple part of Fig.~\ref{fig:model_structure}, the audio-visual branch consists of OC-Softmax integrated with visual and audio extractors, followed by three layers of a feedforward neural network. It is trained with both OC loss and cross-entropy loss. This branch focuses on compacting real-data representations on each feature extractor and separating real- and fake-data representations across both modalities.

The audio-visual model is trained with \(\mathcal{L}_{AV}\): 
\begin{equation}
    \mathcal{L}_{AV} = \mathcal{L}_{AOC}^{av} + \mathcal{L}_{VOC}^{av} + \mathcal{L}_{CE}^{av}
\end{equation}
where 
\(\mathcal{L}_{AOC}^{av}\) and \(\mathcal{L}_{VOC}^{av}\) are the OC-Softmax losses computed using audio and visual features from the audio-visual model with their respective labels.
\(\mathcal{L}_{CE}^{av}\) is the cross-entropy loss applied to the combined audio-visual features after the classifier, using the corresponding 
labels in the audio-visual branch.







\subsection{Inference}
We utilized OC scores, the cosine similarity to the embeddings of bonafide samples of each modality, from both the visual and audio branches. Additionally, we included the AV score, which is the softmax probability of real data from the audio-visual branch. The OC scores were thresholded at 0.5. These thresholded scores were then averaged with the AV score, and a final threshold of 0.5 was applied to determine the prediction.

    
\section{Experimental Setup}

\subsection{Dataset}
The FakeAVCeleb dataset \cite{khalid2021fakeavceleb} includes 500 real videos from various subjects and over 19,500 fake videos. It comprises four categories: RARV (Real Audio, Real Visual, and Synchronized), FARV (Fake Audio, Real Visual), RAFV (Real Audio, Fake Visual), and FAFV (Fake Audio, Fake Visual).

Previous works typically split the FakeAVCeleb dataset by subject ID \cite{zou2024crossmrdf} or randomly \cite{hashmi2022multimodal, muppalla2023integrating}. However, these splits have limitations in assessing the model's generalizability to unseen deepfake generation methods. In this paper, we propose a new split mechanism: we split the dataset based on \textit{generation methods} to evaluate the performance on unseen methods. During the creation of training, validation, and test sets, we ensured that the generation methods used in the test sets were excluded from the training and validation sets.

Our \textbf{Training set} contains 350 videos each from categories RARV (Real), FARV, RAFV, and FAFV (excluding faceswap and faceswap-wav2lip).
\noindent \textbf{Validation set} contains 50 videos each from categories Real, FARV, RAFV, and FAFV (excluding faceswap and faceswap-wav2lip).
\noindent For \textbf{Test set}, we sampled 100 face swap (RAFV) and face swap-wav2lip (FAFV) videos not included in the training and validation sets. We generated 100 audio-only fake videos using a voice conversion library, named category FARV, due to FakeAVCeleb's limited methods for audio fakes. It is important to note that our newly created FARV dataset is synchronized, whereas the FARV from the FakeAVCeleb\cite{khalid2021fakeavceleb} dataset is unsynchronized. Therefore, our FARV dataset has only one cue to detect a fake, making the unseen generation method more distinct. We also created a Unsynced category with 100 unsynchronized videos by shifting audio. Each of these four test datasets — RAFV, FAFV, FARV, and Unsynced— consists of 100 real videos (RARV) and 100 unseen fake videos.

\subsection{Evaluation Measures}

We evaluate audio-visual deepfake detection as a binary classification task based on the final audio-visual label. Accuracy is used as our primary metric, measuring the proportion of correctly classified samples out of the total samples. Given that our four test sets are balanced in terms of real and fake samples, accuracy is an appropriate metric, with a random guess expected to yield close to 50\% accuracy.

\subsection{Comparison Methods}
We adopt some existing methods from the literature for comparison.
The multimodal-dissonance model~\cite{chugh2020dissonance} utilizes a modality dissonance score (a distance between audio and visual features) to detect dissimilarities between modalities. AVDF~\cite{khalid2021evaluation} simply concatenates audio and visual features and maps them directly to audio-visual labels. The multilabel method~\cite{Raza_2023_CVPR} is trained using both audio and visual labels to address the issue that audio-visual labels may confuse the uni-modal feature extractors. MRDF-CE and MRDF-Margin~\cite{zou2024crossmrdf} utilize the cross- and within-modality regularization to maintain the unique features and differences of each modality during multimodal representation learning.

We not only compare our proposed model with state-of-the-art models but also with the  \textbf{A}udio-\textbf{V}isual Branch with \textbf{O}ne-\textbf{C}lass learning (\textbf{AVOC}), the audio-visual branch of MSOC.

\subsection{Training Details}

Our models are trained for 30 epochs using Adam optimizer, with an initial learning rate of $2e-4$ and a batch size of 64. We select the best model with the best Area Under the Curve on the validation set. For the hyperparameter of OC-Softmax, We followed the default parameters from \cite{Zhang2021OC}: $\alpha =20$, $m_0 = 0.9$ and $m_1 = 0.2$. We ran all the models 4 times with different seeds for statistically robust results.

While the three models are trained independently, they share the same training process: training examples are fed to all three models on the same schedule.

Also, for the comparison models, we trained and tested the models in our set-up from scratch for a fair comparison. Specifically, for the multimodal-dissonance model~\cite{chugh2020dissonance}, we trained for 100 epochs.
\begin{table}[]
    \caption{Results of comparison with state-of-the-art models on our test sets derived from the FakeAVCeleb dataset to ensure deepfake generation methods are not seen in training and validation. Average classification accuracy (\%) and standard deviation of four runs of the models are shown.}
    \centering
    \resizebox{0.5\textwidth}{!}{\begin{tabular}{c|cccc}
    \hline
    \textbf{Model} & \textbf{RAFV} & \textbf{FAFV} & \textbf{FARV} & \textbf{Unsynced} \\
    \hline 

Multilabel \cite{Raza_2023_CVPR} & $52.50 \pm 2.50$ & $88.12 \pm 2.19$ & $50.50 \pm 1.80$ & $49.50 \pm 1.62$ \\

    Multimodal-dissonance \cite{chugh2020dissonance} & $48.62 \pm 6.81$ & $62.12 \pm 5.94$ & $57.62 \pm 1.88$ & $49.62 \pm 3.19$ \\
AVDF \cite{khalid2021evaluation} & $50.88 \pm 0.96$ & $86.38 \pm 1.14$ & $51.38 \pm 1.63$ & \textbf{49.88 $\pm$ 2.30} \\
MRDF-CE \cite{zou2024crossmrdf} & $54.38 \pm 2.84$ & $88.25 \pm 0.83$ & $47.25 \pm 0.83$ & $47.50 \pm 0.61$ \\
MRDF-Margin \cite{zou2024crossmrdf} & $55.12 \pm 1.02$ & $86.88 \pm 1.85$ & $47.62 \pm 1.19$ & $47.88 \pm 1.14$ \\

    MSOC (Ours) & \textbf{60.25 $\pm$ 2.19} & \textbf{89.88 $\pm$ 3.15} & \textbf{74.38 $\pm$ 5.41} & $45.25 \pm 1.64$ \\

\hline
    \end{tabular}}
    \label{tab:main_results}
\end{table}

\begin{table}[]
    \caption{The table compares AVOC and MSOC models. Average accuracy (\%) and standard deviation of four runs on each test set. The multilabel model \cite{Raza_2023_CVPR} is used as a baseline.}
    \centering
    \resizebox{0.5\textwidth}{!}{\begin{tabular}{lc|ccccc}
    \hline
    \textbf{Model} &\textbf{Feature Extractor}&  \textbf{RAFV} & \textbf{FAFV} & \textbf{FARV} & \textbf{Unsynced} \\
    \hline 
Multilabel \cite{Raza_2023_CVPR}&- & $52.50 \pm 2.50$ & $88.12 \pm 2.19$ & $50.50 \pm 1.80$ & $49.50 \pm 1.62$ \\
    \hline
    AVOC &SCNet-STIL & \textbf{60.50 $\pm$ 4.06} & $84.38 \pm 2.90$ & $70.62 \pm 1.63$ & $44.12 \pm 1.67$ \\


    MSOC (Ours) &SCNet-STIL & $60.25 \pm 2.19$ & \textbf{89.88 $\pm$ 3.15} & \textbf{74.38 $\pm$ 5.41} & \textbf{45.25 $\pm$ 1.64} \\

\hline
    AVOC &ResNet & $52.75 \pm 2.30$ & $89.12 \pm 4.48$ & $79.62 \pm 2.90$ & \textbf{53.00 $\pm$ 2.89} \\

MSOC &ResNet & \textbf{55.75 $\pm$ 2.02} & \textbf{90.88 $\pm$ 2.43} & \textbf{81.12 $\pm$ 7.45} & $51.75 \pm 2.51$ \\

\hline
    \end{tabular}}
    \label{tab:ms_results}
\end{table}

\section{Results}

\subsection{Comparison with State-of-the-Art Methods}


To demonstrate the effectiveness of the proposed MSOC model on unseen attacks, we compared it with other state-of-the-art audio-visual deepfake detection models.  The comparison of models' performance on test datasets is presented in Table~\ref{tab:main_results}. 
We can observe that state-of-the-art models perform poorly on unseen generation methods, which shows their lack of generalization ability. Our proposed model MSOC outperforms other models on FAFV, RAFV, and FARV test sets.
This indicates that multi-stream architecture with OC-Softmax successfully separated bonafide and generated data by compacting embedding of bonafide data, which resulted in better generalizability than other models in all combinations of fake modality. 

As shown in the last column of Table~\ref{tab:main_results}, all models perform poorly (close to random guessing) when identifying unsynchronized videos, which should be clearly recognized as fake. This is the first time these models have been tested on this unsynchronization benchmark, and our model exhibits general characteristics similar to existing fusion-based methods. The results suggest that training the audio and visual encoders with real/fake labels alone is insufficient to capture synchronization. We believe that incorporating an explicit module to learn audio-visual synchronization~\cite{chung2019perfect, wuerkaixi2022rethinking} could address this issue, but we leave this for future work.

Additionally, we compare the MSOC framework with AVOC models. Table~\ref{tab:ms_results} shows that MSOC models generally perform better than AVOC. This suggests the strength of an audio and visual branch that is only dedicated to separating real and fake in each modality.


\begin{figure}[]
    \centering
    \begin{subfigure}[b]{0.24\textwidth}
        \centering
        \includegraphics[width=\textwidth]{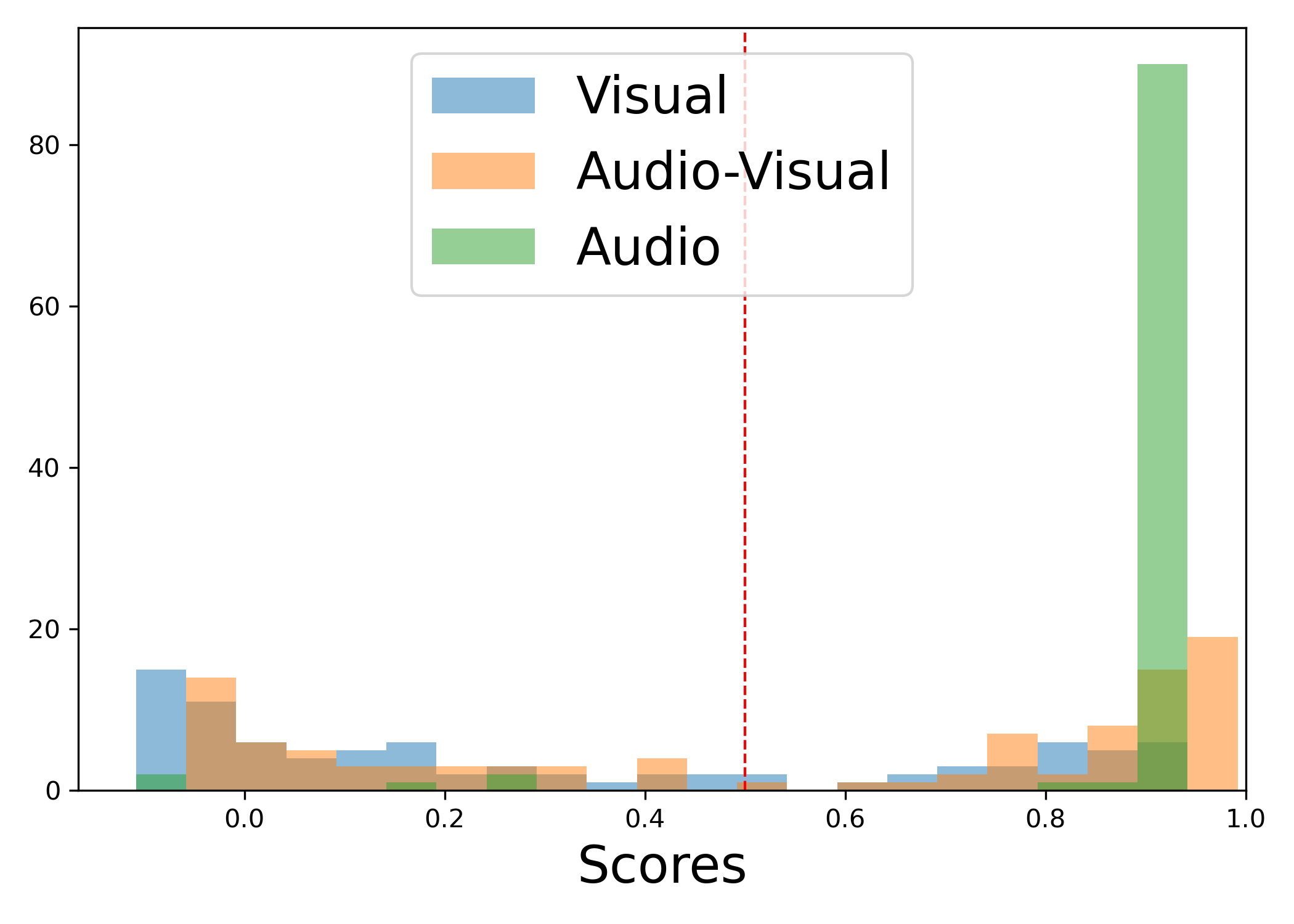}
        \caption{RAFV}
        \label{fig:RAFV}
    \end{subfigure}
    \hfill
    \begin{subfigure}[b]{0.24\textwidth}
        \centering
        \includegraphics[width=\textwidth]{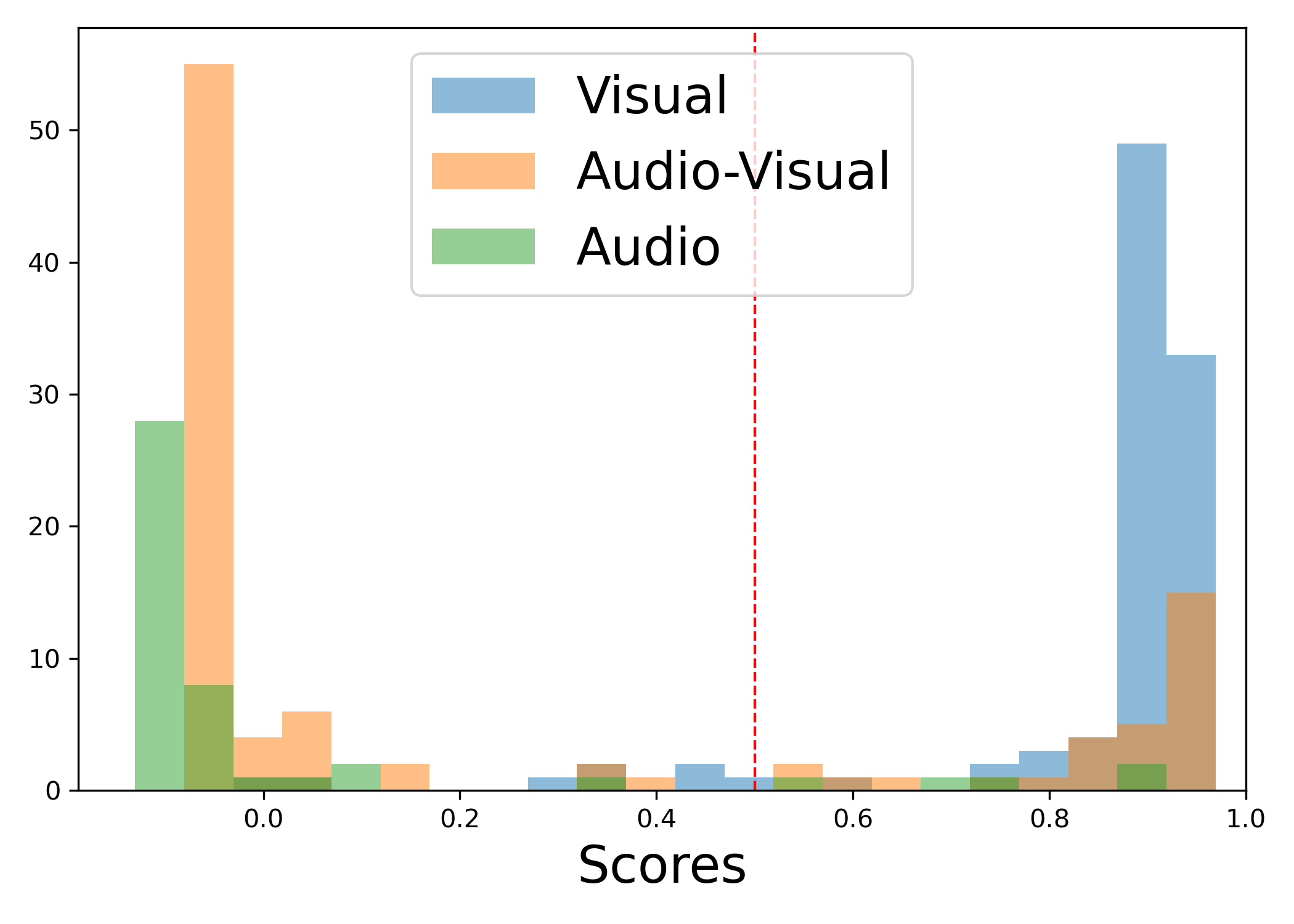}
        \caption{FARV}
        \label{fig:FARV}
    \end{subfigure}
    \\
    \begin{subfigure}[b]{0.24\textwidth}
        \centering
        \includegraphics[width=\textwidth]{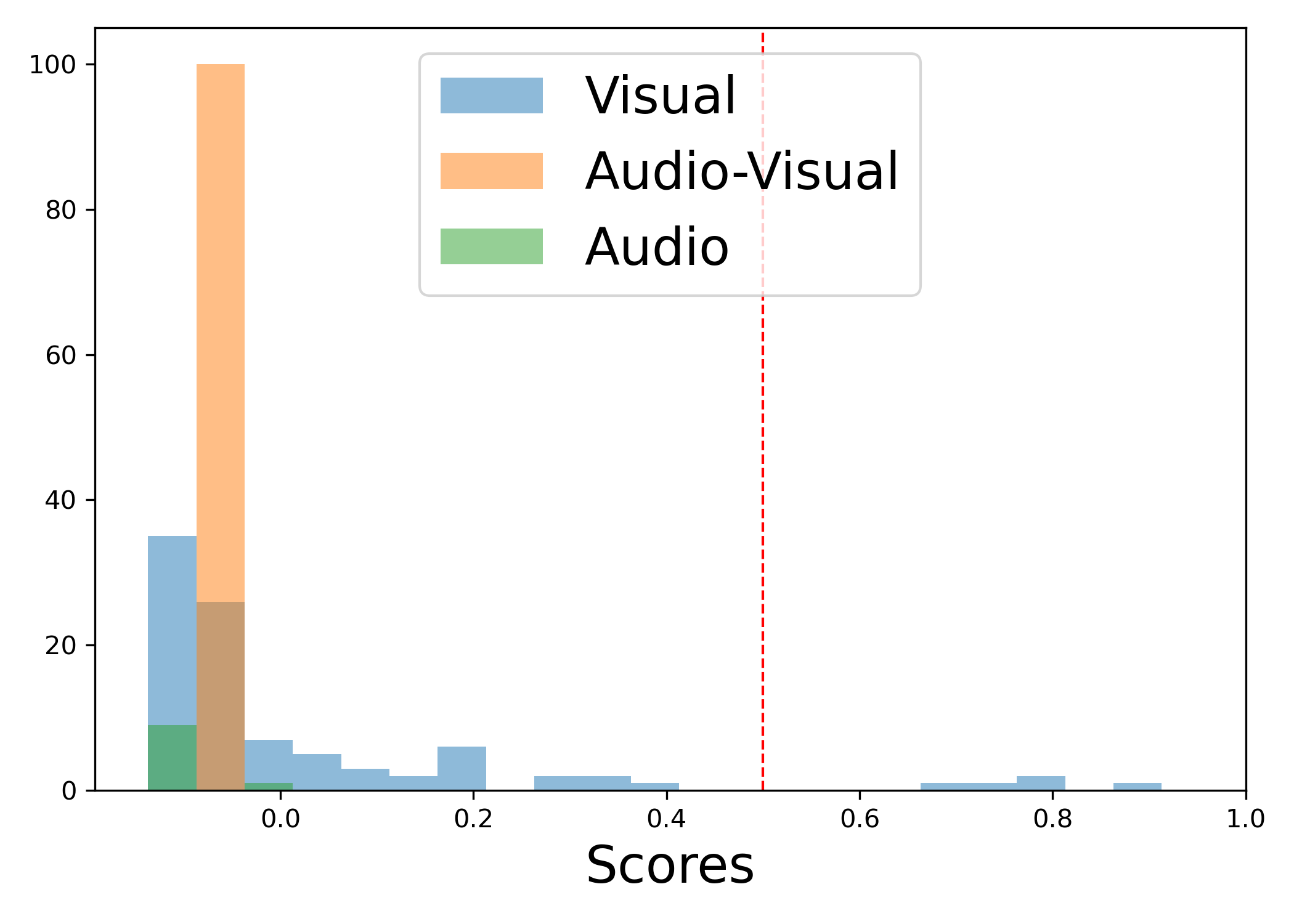}
        \caption{FAFV}
        \label{fig:FAFV}
    \end{subfigure}
    \hfill
    \begin{subfigure}[b]{0.24\textwidth}
        \centering
        \includegraphics[width=\textwidth]{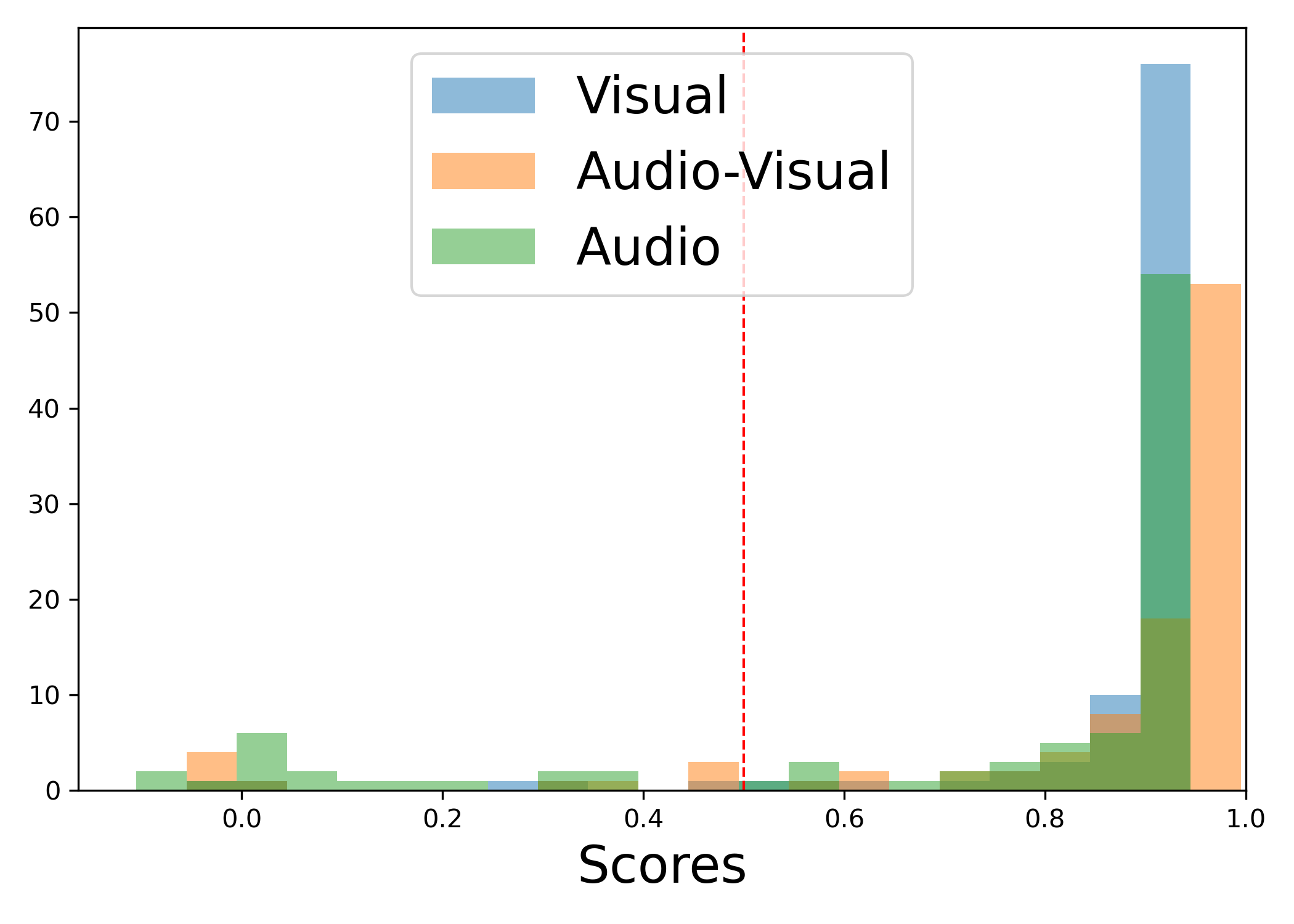}
        \caption{Unsynced}
        \label{fig:unsynced}
    \end{subfigure}
    \caption{Score distribution visualization for four fake categories across each MSOC branch. 
    Scores close to 1 indicate the model perceives the modality as real, and the red vertical line denotes the decision threshold. 
    }
    \label{fig:score_hist}
\end{figure}

\subsection{Performance Analysis of Different Branches of MSOC}
In this section, we delve into the audio and visual branches of the MSOC architecture with the SCNet-STIL visual feature extractor. The MSOC model has three branches, providing enhanced performance and interpretability.

Fig. \ref{fig:score_hist} visualizes the distribution of scores for each branch on four categories of fake videos. 
The audio and visual score, OC scores \(\in [-1, 1]\), are calculated based on the cosine similarity between the bonafide embedding and the feature embedding of the respective modality. The audio-visual score represents the softmax probability that a video is real \(\in [0, 1]\), calculated with audio-visual characteristics and audio-visual labels.
The figure shows that the audio-visual branch performs well when both modalities are fake (FAFV), predicting a probability close to 0 for all fake samples. However, the audio-visual branch exhibits greater confusion when only one modality is fake. The audio branch excels at distinguishing audio fake(Fig.~\ref{fig:FARV}) and real samples(Fig.~\ref{fig:RAFV}). Also, the visual branch exhibits great performance in identifying real samples (Fig.~\ref{fig:FARV}), although it fails to detect some fake samples (Fig.~\ref{fig:RAFV}). This highlights the benefit of using both the audio and visual branches. 

Additionally, the audio and visual branches offer better interpretability of the model's decisions. With AVOC model, it is impossible to determine which modality the model perceives as fake. However, with MSOC, by analyzing the individual scores from branches, one can identify which modality contributes to the final result, providing insights into whether the audio or visual aspect is being manipulated. Therefore, leveraging all branches improves performance and enhances the transparency and reliability of the model's predictions.


\subsection{Impact of One-Class Learning}
In this section, we examine the impact of One-Class Learning by comparing AVOC models trained with and without OC-Softmax. We explore both AVOC models, which are ResNet-based and SCNet-STIL-based. Table~\ref{tab:oc_results} shows that the AVOC models trained with the OC-Softmax generally outperform AVOC models trained without the guidance of OC-Softmax. This result exhibits that implementing one-class learning on audio-visual deepfake detection successfully enhances models' robustness to unseen attacks by compacting the bonafide representations.

\begin{table}[]
    \centering
    \caption{Comparison of models trained with and without OC softmax using different feature extractors. Average accuracy (\%) and standard deviation of four runs. The multilabel model \cite{Raza_2023_CVPR} is used as a baseline.}
    \resizebox{0.5\textwidth}{!}{\begin{tabular}{lcc|cccc}
        \hline
        \textbf{Model} & \textbf{Feature Extractor} & \textbf{OC} & \textbf{RAFV} & \textbf{FAFV} & \textbf{FARV} & \textbf{Unsynced} \\

        \hline
Multilabel \cite{Raza_2023_CVPR}&-&- & $52.50 \pm 2.50$ & $88.12 \pm 2.19$ & $50.50 \pm 1.80$ & $49.50 \pm 1.62$ \\

        \hline
        AVOC & SCNet-STIL & No & $50.12 \pm 2.01$ & $77.50 \pm 1.17$ & $68.75 \pm 6.03$ & \textbf{48.88 $\pm$ 1.98} \\
        AVOC & SCNet-STIL & Yes & \textbf{60.50 $\pm$ 4.06} & \textbf{84.38 $\pm$ 2.90} & \textbf{70.62 $\pm$ 1.63} & $44.12 \pm 1.67$ \\
        \hline
        AVOC & ResNet & No & $46.75 \pm 3.78$ & $74.00 \pm 0.94$ & $69.00 \pm 5.18$ & $45.00 \pm 2.47$ \\
        AVOC & ResNet & Yes & \textbf{52.75 $\pm$ 2.30} & \textbf{89.12 $\pm$ 4.48} & \textbf{79.62 $\pm$ 2.90} & \textbf{53.00 $\pm$ 2.89} \\
        \hline
    \end{tabular}}
    \label{tab:oc_results}
\end{table}

\begin{figure}[]
    \centering
    \includegraphics[width=0.5\textwidth]{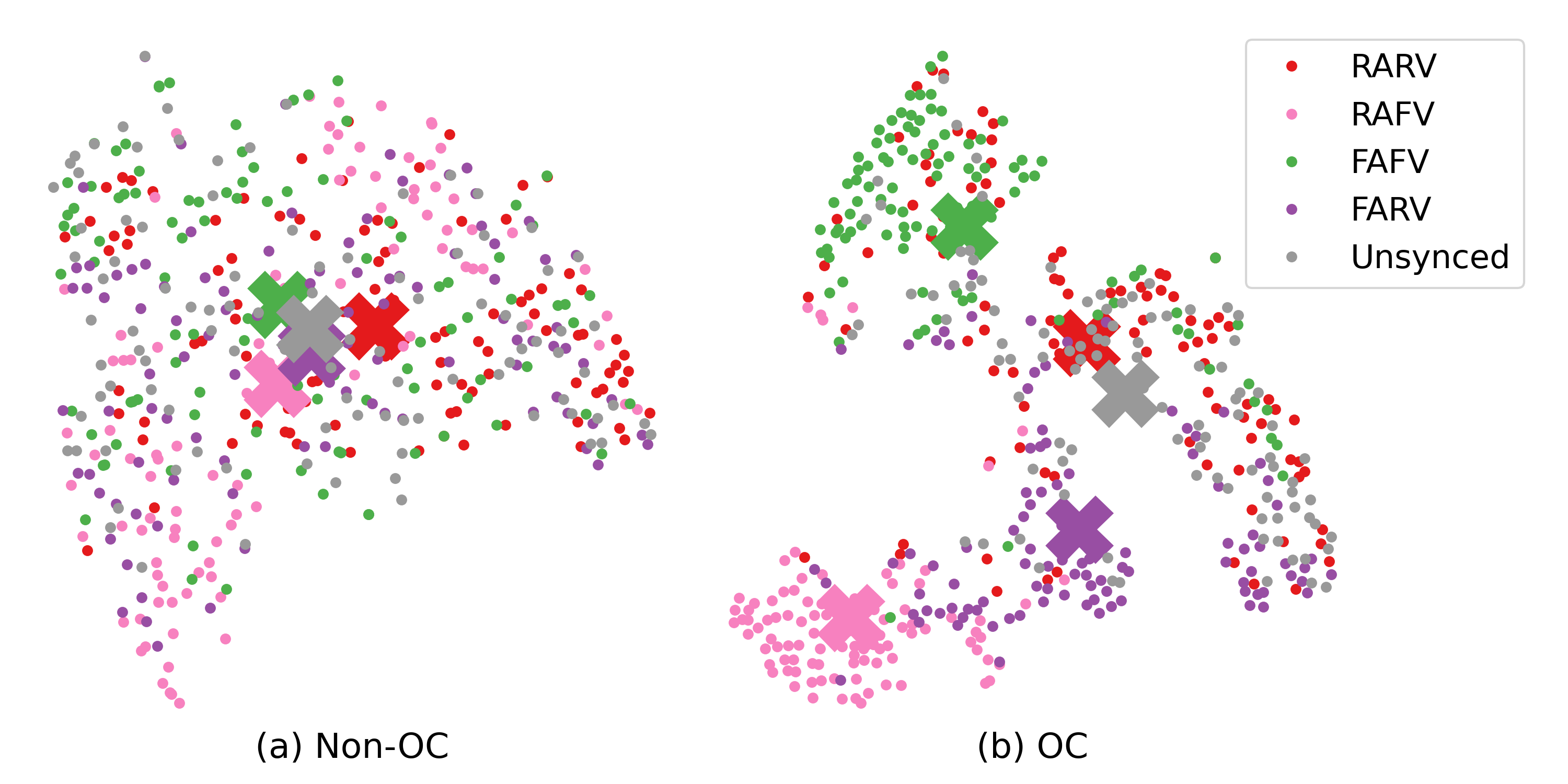}
    \caption{t-SNE visualization of concatenated audio-visual feature. The cross ``X'' in the figure represents the center of the data for each category. Better viewed in color.}
    \label{fig:tsne}
\end{figure}

We visualized the impact of OC-Softmax in Fig. \ref{fig:tsne}
by comparing the audio-visual embeddings of the model trained with and without OC-Softmax. The model trained with OC-Softmax successfully separates fake categories RAFV, FAFV, and FARV from real samples (RARV), although Unsynchronized samples still exhibit some overlap with the real samples. This overlap is anticipated, as detecting the unsynchronization is beyond the scope of an uni-modal feature extractor.



\subsection{Impact of Visual Feature Extractor}
Table~\ref{tab:ms_results} demonstrates that models with SCNet-STIL visual feature extractor perform better on the RAFV test set. Thus, 
this section examines the impact of visual feature extractors in one-class learning. Although OC-Softmax effectively compacts genuine representations and distributes fake representations, its performance is limited if the visual feature extractor fails to capture the general features of fake visual artifacts. This limitation arises because OC-Softmax compacts real representations based on observed attacks and real data, potentially including unseen attack representations within the realm of genuine representations. Therefore, extracting more general features of fake videos, such as Spatial-Temporal Inconsistency, could be beneficial. 

\begin{figure}[t]
    \centering
    \begin{subfigure}[b]{0.24\textwidth}
        \centering
        \includegraphics[width=\textwidth]{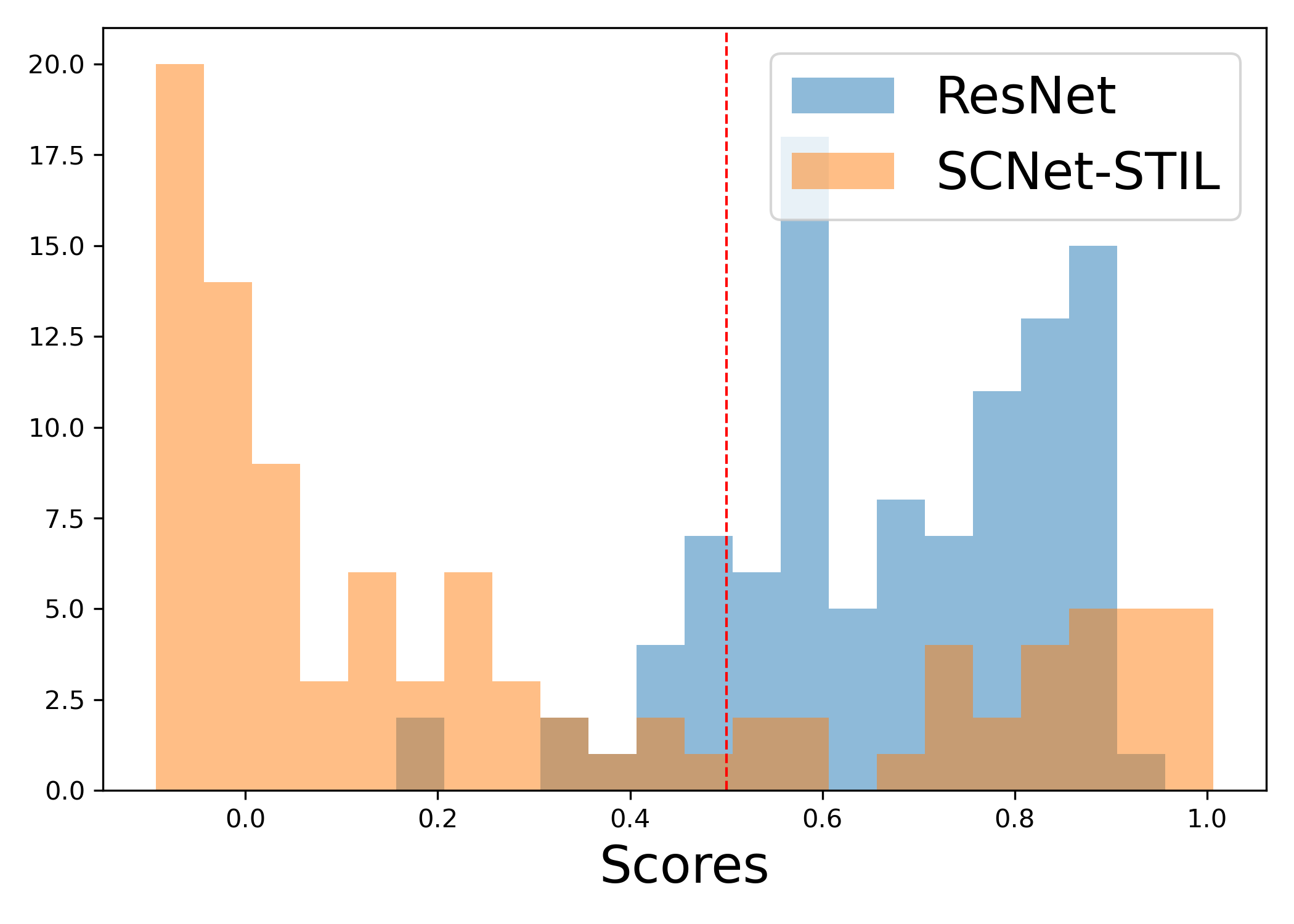}
        \caption{Fake}
        \label{fig:Fake}
    \end{subfigure}
    \hfill
    \begin{subfigure}[b]{0.24\textwidth}
        \centering
        \includegraphics[width=\textwidth]{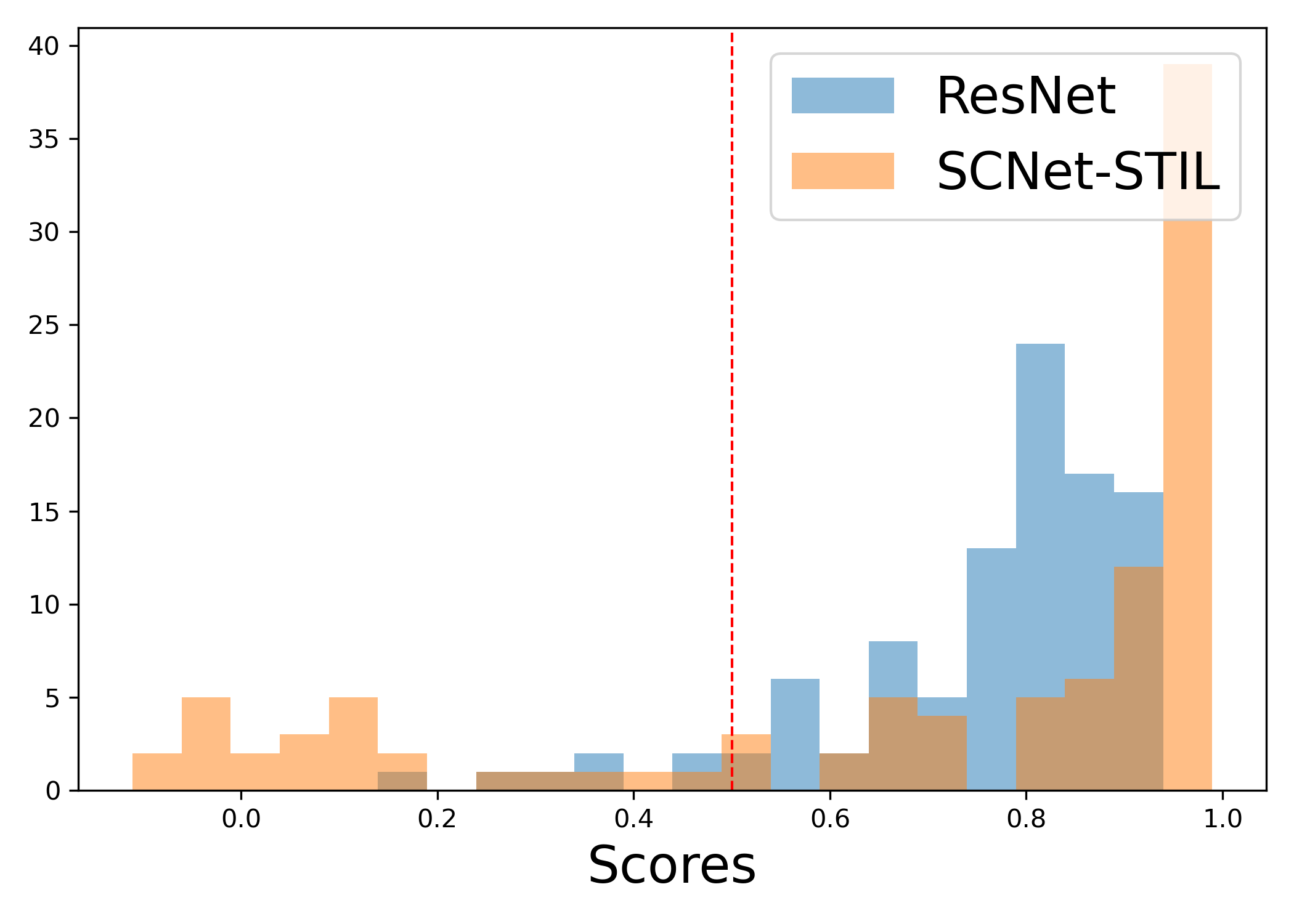}
        \caption{Real}
        \label{fig:Real}
    \end{subfigure}

    \caption{The figure compares visual scores computed from ResNet and SCNet-STIL visual feature extractors on both fake and real samples.}
    \label{fig:compare_vfs}
\end{figure}

Fig.~\ref{fig:compare_vfs} compares the visual scores from both visual feature extractors. We can observe that the ResNet-based visual feature extractor lacks the ability to detect unseen fake methods effectively compared to the SCNet-STIL-based visual feature extractor. This explains why models with the STIL feature extractor significantly outperform models with a ResNet feature extractor on the RAFV test set.


\section{Conclusion}
This paper presents a multi-stream fusion framework with one-class learning to enhance audio-visual deepfake detection. Our proposed framework improves detection performance against unseen deepfake generation methods compared to SOTA models.
Additionally, the MSOC framework provides interpretability, offering the ability to identify which modality is fake, which can be achieved through the score distribution of the models (Audio, Visual, Audio-Visual). Future work includes joint modeling of detecting audio-visual unsynchronization and deepfakes and a more robust framework for rooting the fake modality.

\bibliographystyle{IEEEtran}
\bibliography{main}

\begin{thebibliography}{10}
\providecommand{\url}[1]{#1}
\csname url@samestyle\endcsname
\providecommand{\newblock}{\relax}
\providecommand{\bibinfo}[2]{#2}
\providecommand{\BIBentrySTDinterwordspacing}{\spaceskip=0pt\relax}
\providecommand{\BIBentryALTinterwordstretchfactor}{4}
\providecommand{\BIBentryALTinterwordspacing}{\spaceskip=\fontdimen2\font plus
\BIBentryALTinterwordstretchfactor\fontdimen3\font minus \fontdimen4\font\relax}
\providecommand{\BIBforeignlanguage}[2]{{%
\expandafter\ifx\csname l@#1\endcsname\relax
\typeout{** WARNING: IEEEtran.bst: No hyphenation pattern has been}%
\typeout{** loaded for the language `#1'. Using the pattern for}%
\typeout{** the default language instead.}%
\else
\language=\csname l@#1\endcsname
\fi
#2}}
\providecommand{\BIBdecl}{\relax}
\BIBdecl

\bibitem{rombach2022stablediff}
R.~Rombach, A.~Blattmann, D.~Lorenz, P.~Esser, and B.~Ommer, ``High-resolution image synthesis with latent diffusion models,'' in \emph{Proc. CVPR}, 2022.

\bibitem{tan2021survey}
X.~Tan, T.~Qin, F.~Soong, and T.-Y. Liu, ``A survey on neural speech synthesis,'' \emph{arXiv preprint arXiv:2106.15561}, 2021.

\bibitem{desai2009voiceconversion}
S.~Desai, E.~V. Raghavendra, B.~Yegnanarayana, A.~W. Black, and K.~Prahallad, ``Voice conversion using artificial neural networks,'' in \emph{Proc. ICASSP}.\hskip 1em plus 0.5em minus 0.4em\relax IEEE, 2009.

\bibitem{sisman2020overview}
B.~Sisman, J.~Yamagishi, S.~King, and H.~Li, ``An overview of voice conversion and its challenges: From statistical modeling to deep learning,'' \emph{IEEE/ACM TASLP}, vol.~29, 2020.

\bibitem{Korshunova2017faceswap}
I.~Korshunova, W.~Shi, J.~Dambre, and L.~Theis, ``Fast face-swap using convolutional neural networks,'' in \emph{Proc. ICCV}, Oct 2017.

\bibitem{sheng2024deep}
C.~Sheng, G.~Kuang, L.~Bai, C.~Hou, Y.~Guo, X.~Xu, M.~Pietik{\"a}inen, and L.~Liu, ``Deep learning for visual speech analysis: A survey,'' \emph{IEEE TPAMI}, 2024.

\bibitem{prajwal2020lipsync}
K.~Prajwal, R.~Mukhopadhyay, V.~P. Namboodiri, and C.~Jawahar, ``A lip sync expert is all you need for speech to lip generation in the wild,'' in \emph{Proc. ACM MM}, 2020.

\bibitem{guan2023stylelipsync}
J.~Guan, Z.~Zhang, H.~Zhou, T.~Hu, K.~Wang, D.~He, H.~Feng, J.~Liu, E.~Ding, Z.~Liu \emph{et~al.}, ``Stylesync: High-fidelity generalized and personalized lip sync in style-based generator,'' in \emph{Proc. CVPR}, 2023.

\bibitem{zou2024crossmrdf}
H.~Zou, M.~Shen, Y.~Hu, C.~Chen, E.~S. Chng, and D.~Rajan, ``Cross-modality and within-modality regularization for audio-visual deepfake detection,'' in \emph{Proc. ICASSP}, 2024.

\bibitem{muppalla2023integrating}
S.~Muppalla, S.~Jia, and S.~Lyu, ``Integrating audio-visual features for multimodal deepfake detection,'' \emph{arXiv preprint arXiv:2310.03827}, 2023.

\bibitem{zhou2021joint}
Y.~Zhou and S.-N. Lim, ``Joint audio-visual deepfake detection,'' in \emph{Proc. ICCV}, 2021.

\bibitem{khalid2021fakeavceleb}
H.~Khalid, S.~Tariq, M.~Kim, and S.~S. Woo, ``Fake{AVC}eleb: A novel audio-video multimodal deepfake dataset,'' in \emph{Proc. NeurIPS Datasets and Benchmarks Track}, 2021.

\bibitem{dolhansky2020dfdc}
B.~Dolhansky, J.~Bitton, B.~Pflaum, J.~Lu, R.~Howes, M.~Wang, and C.~C. Ferrer, ``The deepfake detection challenge ({DFDC}) dataset,'' \emph{arXiv preprint arXiv:2006.07397}, 2020.

\bibitem{Zhang2021OC}
Y.~Zhang, F.~Jiang, and Z.~Duan, ``One-class learning towards synthetic voice spoofing detection,'' \emph{IEEE Signal Processing Letters}, vol.~28, 2021.

\bibitem{lu2024onedis}
J.~Lu, Y.~Zhang, W.~Wang, Z.~Shang, and P.~Zhang, ``One-class knowledge distillation for spoofing speech detection,'' in \emph{Proc. ICASSP}, 2024.

\bibitem{hu2022finfervideo}
J.~Hu, X.~Liao, J.~Liang, W.~Zhou, and Z.~Qin, ``Finfer: Frame inference-based deepfake detection for high-visual-quality videos,'' in \emph{Proc. AAAI}, vol.~36, no.~1, 2022.

\bibitem{shahzad2022lip}
S.~A. Shahzad, A.~Hashmi, S.~Khan, Y.-T. Peng, Y.~Tsao, and H.-M. Wang, ``Lip sync matters: A novel multimodal forgery detector,'' in \emph{Proc. APSIPA ASC}.\hskip 1em plus 0.5em minus 0.4em\relax IEEE, 2022.

\bibitem{chugh2020dissonance}
K.~Chugh, P.~Gupta, A.~Dhall, and R.~Subramanian, ``Not made for each other-audio-visual dissonance-based deepfake detection and localization,'' in \emph{Proc. ACM MM}, 2020.

\bibitem{yang2023avoiddf}
W.~Yang, X.~Zhou, Z.~Chen, B.~Guo, Z.~Ba, Z.~Xia, X.~Cao, and K.~Ren, ``{AVoiD-DF}: Audio-visual joint learning for detecting deepfake,'' \emph{IEEE TIFS}, vol.~18, 2023.

\bibitem{hashmi2023avtenet}
A.~Hashmi, S.~A. Shahzad, C.-W. Lin, Y.~Tsao, and H.-M. Wang, ``Avtenet: Audio-visual transformer-based ensemble network exploiting multiple experts for video deepfake detection,'' \emph{arXiv preprint arXiv:2310.13103}, 2023.

\bibitem{hashmi2022multimodal}
A.~Hashmi, S.~A. Shahzad, W.~Ahmad, C.~W. Lin, Y.~Tsao, and H.-M. Wang, ``Multimodal forgery detection using ensemble learning,'' in \emph{Proc. APSIPA ASC}.\hskip 1em plus 0.5em minus 0.4em\relax IEEE, 2022.

\bibitem{asha2022novel}
S.~Asha, P.~Vinod, I.~Amerini, and V.~G. Menon, ``A novel deepfake detection framework using audio-video-textual features,'' 2022.

\bibitem{Ivanovska2021dis_oc}
M.~Ivanovska and V.~{\v{S}}truc, ``A comparative study on discriminative and one--class learning models for deepfake detection,'' 2021.

\bibitem{ding2023samo}
S.~Ding, Y.~Zhang, and Z.~Duan, ``{SAMO}: Speaker attractor multi-center one-class learning for voice anti-spoofing,'' in \emph{Proc. ICASSP}, 2023.

\bibitem{kim2024one}
H.~M. Kim, K.~Jang, and H.~Kim, ``One-class learning with adaptive centroid shift for audio deepfake detection,'' in \emph{Proc. Interspeech}, 2024.

\bibitem{he2016deep}
K.~He, X.~Zhang, S.~Ren, and J.~Sun, ``Deep residual learning for image recognition,'' in \emph{Proc. CVPR}, 2016.

\bibitem{liu2020scnet}
J.-J. Liu, Q.~Hou, M.-M. Cheng, C.~Wang, and J.~Feng, ``Improving convolutional networks with self-calibrated convolutions,'' in \emph{Proc. CVPR}, 2020.

\bibitem{gu2021spatiotemporal}
Z.~Gu, Y.~Chen, T.~Yao, S.~Ding, J.~Li, F.~Huang, and L.~Ma, ``Spatiotemporal inconsistency learning for deepfake video detection,'' in \emph{Proc. ACM MM}, 2021.

\bibitem{khalid2021evaluation}
H.~Khalid, M.~Kim, S.~Tariq, and S.~S. Woo, ``Evaluation of an audio-video multimodal deepfake dataset using unimodal and multimodal detectors,'' in \emph{Proc. ACM MM ADGD Workshop}, 2021.

\bibitem{Raza_2023_CVPR}
M.~A. Raza and K.~M. Malik, ``Multimodaltrace: Deepfake detection using audiovisual representation learning,'' in \emph{Proc. CVPR Workshops}, 2023.

\bibitem{chung2019perfect}
S.-W. Chung, J.~S. Chung, and H.-G. Kang, ``Perfect match: Improved cross-modal embeddings for audio-visual synchronisation,'' in \emph{Proc. ICASSP}.\hskip 1em plus 0.5em minus 0.4em\relax IEEE, 2019, pp. 3965--3969.

\bibitem{wuerkaixi2022rethinking}
A.~Wuerkaixi, Y.~Zhang, Z.~Duan, and C.~Zhang, ``Rethinking audio-visual synchronization for active speaker detection,'' in \emph{Proc. IEEE MLSP}, 2022.

\end{thebibliography}
\end{document}